\documentclass[reprint,amsmath,amssymb,aps,pra]{revtex4}

\usepackage{graphicx}
\usepackage{dcolumn}
\usepackage{bm}
\usepackage{amsthm}
\theoremstyle{plain}
\newtheorem*{theorem*}{Theorem}

\newcommand{\C}{\mathbb{C}}
\newcommand{\R}{\mathbb{R}}

\newcommand{\rank}{\text{rank}}

\begin{document}

\preprint{APS/123-QED}

\title{Entanglement on Orbits}

\author{Songbo Xie}
\email{sxie9@ur.rochester.edu}
\affiliation{Center for Coherence and Quantum Optics, and Department of Physics and Astronomy, University of Rochester, Rochester NY 14627 USA.
}


\begin{abstract}
The study of multipartite entanglement is not only interesting but also important due to its wide application in quantum information processing. However, the complicated structure of the Hilbert space for many parties makes multipartite entanglement extremely complicated. It is then worth studying the structure of the Hilbert space itself. In this work, we provide a way to study the structure of SLOCC-equivalence and to determine the number free parameters for SLOCC-equivalent classes. Additionally, two different entanglement witnesses are introduced. The method matches well the existing results, and can make predictions for more-qubit systems. 
\end{abstract}

\maketitle

\section{Introduction}
Entanglement, especially multipartite entanglement, is an ongoing topic not only conceptually interesting, but also practically significant due to the recognition of entanglement as a useful resource in quantum information \cite{chitambar2019quantum}. In quantum resource theory, entanglement is not only a qualitative concept, but is also quantified and seen as a common resource shared between multiple parties, where each party possesses one part of an overall entangled quantum system. Among all cases, the qubit systems, where each party has only two levels, are of great interests to quantum information and quantum computation due to their powerful simplicity compared to higher-dimensional systems, as well as the strong similarity between an n-qubit system and an n-bit traditional computing system. 

The quantification of entanglement for two-qubit systems has been studied since decades ago, which turns out to be trivial in the sense that all two-qubit entanglement measures are ``equivalent''---they always agree on the relative rankings of entanglement amount between any two states. However, the multi-qubit situation is rather complicated. One of the most significant reasons is the extremely high dimensionality of the Hilbert space for multi-qubit systems, which grants the possibility of having inequivalent multipartite entanglement measures (see Vidal \cite{vidal2000entanglement}). The complication of multi-qubit Hilbert space implies that we should divert attention from entanglement to the study of multi-qubit Hilbert space itself.

One of the routes to attack this problem is to simplify the structure of the Hilbert space by equivalent relations. Specifically, entanglement measures have an important property called {\it local monotonicity}, that is, entanglement should be nonincreasing when a state is converted into another one by local quantum operations assisted with classical communications (LOCC). When two states are mutually convertible into each other with certainty under LOCC, it follows that the two states have the same amount of entanglement \cite{plenio1998teleportation}. Mathematically, this requires a local unitary (LU) operation between the two states $|\psi\rangle$ and $|\phi\rangle$ such that $|\psi\rangle=U_1\otimes\cdots\otimes U_n|\phi\rangle$, where $U_i$ with $i=1,\cdots, n$ are 2 by 2 unitary matrices. For this reason, the two quantum states are said to be of {\it LU-equivalence}. It implies that entanglement is invariant under local basis transformations.

LU-equivalence, however, does not simplify the Hilbert space to a satisfying extent. Even in the simplest two-qubit case, the whole Hilbert space is identified as one-dimensional, and a continuous variable is still needed, \textit{e.g.}, the angle $\theta$ in the Schmidt decomposition \cite{ekert1995entangled}
\begin{equation}
    |\psi\rangle=\cos\theta|00\rangle+\sin\theta|11\rangle,\quad 0\leq\theta\leq\pi/4.
\end{equation}
Here, each $\theta$ value represents an LU-equivalent class, and different $\theta$ values have different entanglements. For the three-qubit case, five such free parameters are needed according to Ac\'in {\it et al}. \cite{acin2000generalized}. Hence we still need to deal with infinitely many different equivalent classes. 

There is, however, another kind of equivalent relation, where two states are called {\it stochastically equivalent} when the conversion rate under LOCC from one state to the other one is nonvanishing \cite{dur2000three}. This kind of relation is called SLOCC-equivalence (SLOCC is the abbreviation for stochastic LOCC). Mathematically, this requires an invertible local operation between the two states $|\psi\rangle$ and $|\phi\rangle$ such that $|\psi\rangle=A_1\otimes\cdots\otimes A_n|\phi\rangle$, where $A_i$ with $i=1,\cdots, n$ are 2 by 2 invertible matrices. Since the number of invertible matrices is much more than that of unitary matrices, the Hilbert space under SLOCC-equivalence turns out to be much simpler than LU-equivalence. 

SLOCC-equivalence was studied for three-qubit systems \cite{dur2000three}. There are only six distinct SLOCC-equivalent classes, of which only two are ``genuinely'' entangled, namely the GHZ class and the $W$ class. The GHZ class has five ``free parameters'' (the meaning of free parameters shall be explained later), and is dense and occupies almost everywhere in the Hilbert space. The $W$ class has only three free parameters and is much smaller compared to the GHZ class. A few tripartite entanglement measures have been studied based on the SLOCC-equivalence classification (examples are \cite{xie2021triangle,ma2011measure}).

It is intriguing to follow the SLOCC-equivalence procedure in more-qubit systems. Attempts for four qubits have been made and it was found by Verstraete {\it et al.} that the number of ``families'' in four-qubit systems is nine \cite{verstraete2002four}. It is easy to check that their defined ``family'' is not the SLOCC-equivalent class. In contrast, Li {\it et al.} \cite{li2007classification} identified the presence of at least 28 distinct SLOCC equivalent classes in four-qubit systems. Based on this information, it is thus urgent to develop a mathematical tool to determine the structure of SLOCC-equivalent classes for systems containing an arbitrary number of qubits. 

In \cite{linden1998multi}, Linden and Popescu brought the idea of orbits from Lie groups and advanced techniques to determine the dimension and ``invariants'' of LU-equivalent classes in the Hilbert space $\C^{2^n}$ for arbitrary n qubits. Lyons and Walck \cite{lyons2005minimum,lyons2006classification,walck2007maximum,lyons2008classification} developed the techniques to study LU-equivalence in the ``state space'' $\C P^{2^n-1}$, where the states are insensitive to an overall complex factor. In this work, we are inspired by these previous techniques and develop them to further study the structure of SLOCC-equivalent classes in four-qubit systems. It then turns out to be a powerful tool to study multipartite entanglement measures. A method to determine the dimension of the SLOCC quotient space for arbitrary numbers of qubits is also advanced. It matches well the already existing results, and can make predictions in more-qubit cases. What's more, an entanglement witness is introduced together along, which can detect different entanglement types. However, different previous literature mixes their uses of the group U(2) and SU(2), without explaining well their distinction. We argue that the uses of U(2) and SU(2), or in our case of SL(2,$\C$) and GL(2,$\C$), are indeed different for orbits in the ket space, for which the distinction can be seen as a special entanglement witness, which can detect the GHZ-type genuine entanglement.

\section{Notations}
A similar discussion of the notations can be found in \cite{lyons2005minimum}.\\

\noindent{\bf Hilbert space and state space.}\quad The Hilbert space $H$ for an n-qubit pure-state system is composed of ``kets'' in the Dirac notation, where each ket is expressed with $2^n$ complex numbers as
\begin{equation}\label{ket}
    |\psi\rangle=\sum_{i_1,\cdots,i_n=0}^1c_{i_1\cdots i_n}|i_1\cdots i_n\rangle.
\end{equation}
Here, each $i_k$ takes a value of either 0 or 1, which determines the $k$th qubit. Therefore, the Hilbert space is also called ket space. Due to the relation, $H\simeq\C^{2^n}\simeq\R^{2^{n+1}}$, one can identify an arbitrary ket as a complex $2^n$-tuple or a real $2^{n+1}$-tuple in the Euclidean space. For our notation,
\begin{equation}
    |\psi\rangle\longmapsto\begin{pmatrix}
    a_{0\cdots 0}+ib_{0\cdots 0}\\
    a_{0\cdots 1}+ib_{0\cdots 1}\\
    \vdots\\
    a_{1\cdots 1}+ib_{1\cdots 1}
    \end{pmatrix}\longmapsto
    \begin{pmatrix}
    a_{0\cdots 0}\\
    b_{0\cdots 0}\\
    a_{0\cdots 1}\\
    b_{0\cdots 1}\\
    \vdots\\
    a_{1\cdots 1}\\
    b_{1\cdots 1}
    \end{pmatrix},
\end{equation}
where $a_{i_1\cdots i_n}=\Re(c_{i_1\cdots i_n})$ and $b_{i_1\cdots i_n}=\Im(c_{i_1\cdots i_n})$ are the real and imaginary parts respectively.

A state is slightly different from a ket. As Dirac noted in \cite{dirac1981principles}, the superposition of two kets, $c_1|\psi\rangle$ and $c_2|\psi\rangle$, which is written as $(c_1+c_2)|\psi\rangle$, must corresponds to the same state as $|\psi\rangle$ does. Equivalently speaking, two kets that differ by an overall complex factor represent the same state. A quantum state is a ray in the Hilbert space---$\{c|\psi\rangle\}$, where $c$ is any nonzero complex number. If we recognize the following equivalent class
\begin{equation}
    \sim\ :\ (c_1,\cdots,c_{2^n})\equiv(\lambda c_1,\cdots,\lambda c_{2^n}),\ \forall\lambda\in\C,
\end{equation}
the state space is then the complex projective space $HP=\mathbb{C}P^{2^n-1}=\left.\C^{2^n}\right/\sim$, where the equivalent class $\sim$ is defined above. In the state space, neither the real normalization factor nor the $U(1)$ global phase plays a role. 

We denote a quantum state by a point $x$ in the state space $HP$. If the ket $\psi\in H$ corresponds to the state $x\in HP$, we say $\psi$ is a representative of $x$, and denote it as $x=[\psi]$. It is obvious that $[\psi]=[\lambda\psi],\ \forall\lambda\in\C$. A fibration map from the ket space to the state space can be visualized as:
\begin{equation}
    \begin{split}
        \pi:\ H&\longrightarrow HP,\\
        \pi:\ \psi&\longmapsto [\psi].
    \end{split}
\end{equation}

\section{Orbits in Ket and State Spaces}

The SLOCC-equivalence involves a group acting on a space. What is the group and what is the space?

As discussed above, the ket space cares about the overall complex factor, whereas the state space does not. Similarly, there are also two groups, GL(2,$\C$)$^{\otimes n}$ and SL(2,$\C$)$^{\otimes n}$.

According to D\"ur, Vidal, and Cirac \cite{dur2000three}, two kets $\psi,\phi\in H$ are called stochastically equivalent (SLOCC-equivalence) if they have a nonvanishing probability of success when trying to convert $\psi$ into $\phi$ under LOCC. Equivalently, as they have mentioned, there exists a matrix $g$ connecting the two kets as $\psi=g\phi$, where $g$ is of the form $A_1\otimes A_2\otimes\cdots\otimes A_n$, where all the $A$ matrices are invertible. That is, $\psi$ and $\phi$ are related by an ``invertible local operator (ILO)''. Mathematically, the action $g$ is in the group GL(2,$\C$)$^{\otimes n}$. However, a few works following that path use instead the group SL(2,$\C$)$^{\otimes n}$. An element in SL(2,$\C$)$^{\otimes n}$, is of the form $B_1\otimes B_2\otimes\cdots\otimes B_n$, where all the $B$ matrices are not only invertible, but also with determinant 1. Their consideration is that physical state is insensitive to local phases and local normalizations. So why don't we simply take them out from the very beginning in the group actions. The consideration is definitely correct intuitively. However, we are also interested in the ket space, for which global phases and normalization do matter. And the results for the actions of the two groups GL(2,$\C$)$^{\otimes n}$ and SL(2,$\C$)$^{\otimes n}$ are indeed different on the ket space. And it turns out that the difference is not trivial, which can even reveal a special entanglement feature. In the following, we shall discuss the actions of the two distinct groups on the two distinct spaces respectively.

\subsection{Orbits in ket space}

A smooth left action of the group $G$ on the ket space $H$ is naturally defined as the smooth map:
\begin{equation}
\begin{split}
    &L:G\times H\longrightarrow H, \\
    &L:g\times \psi\longmapsto g\psi\ s.t.\\
    (a)\ &\forall g\in G,\ L_g:H\rightarrow H\ \text{is a diffeomorphism}.\\
    (b)\ &\forall g,h\in G,\ L_{gh}=L_g\circ L_h.
\end{split}
\end{equation}
We first take the group $G$ as $\text{GL}(2,\C)^{\otimes n}$. An element $g\in G$ is a complex $2^n$ by $2^n$ matrix. The multiplication $g\psi$ here is the usual matrix multiplication by a column vector. 

\begin{figure}[b]
    \centering
    \includegraphics[width=0.6\textwidth]{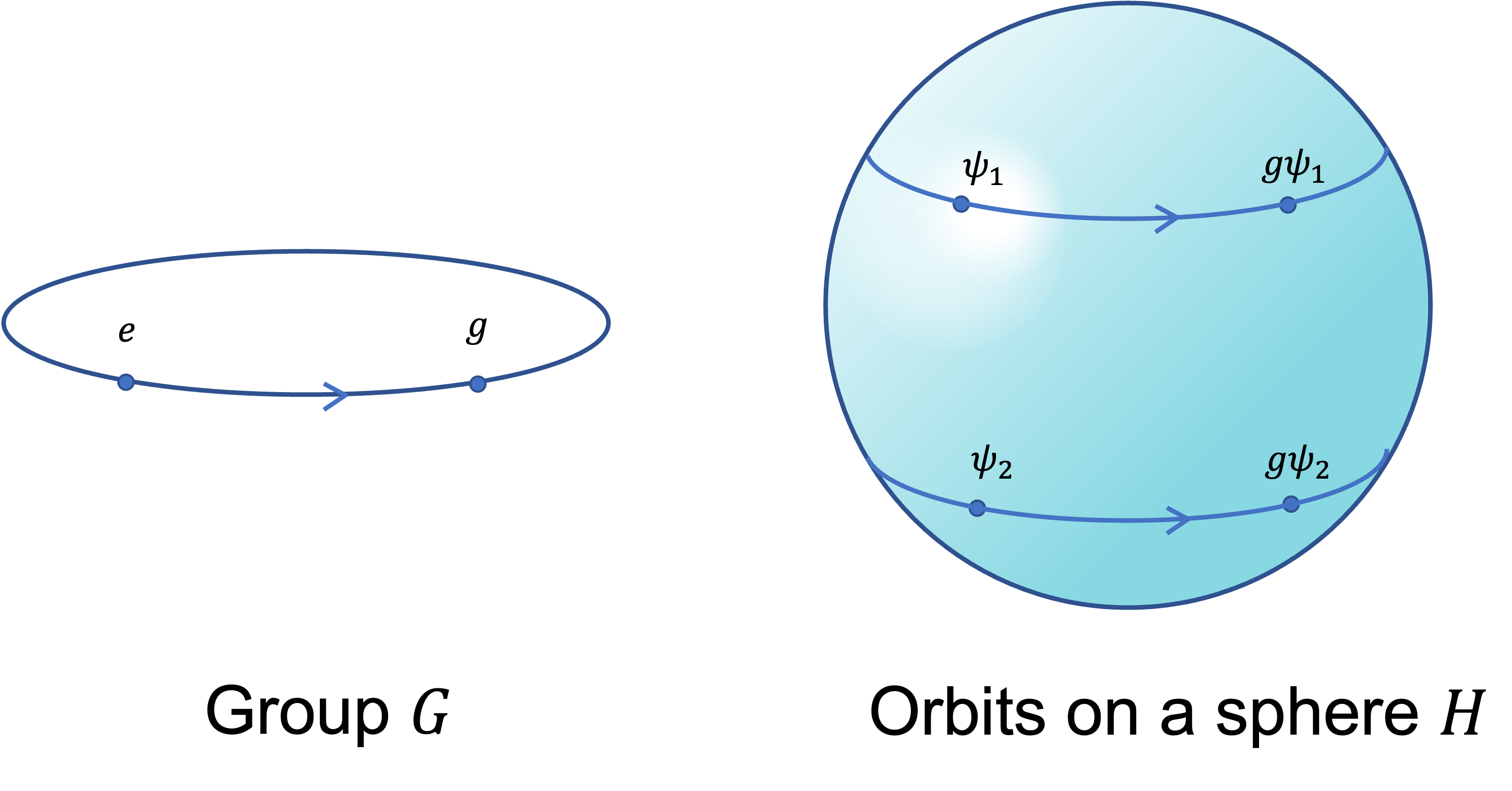}
    \caption{The group $G$ has an action on the space $H$: $G\times H\rightarrow H$. Given a state $\psi_1\in H$, there is a map from the group $G$ to the orbit of $\psi_1$. Given a state $\psi_2\in H$, there is a map from the group $G$ to the orbit of $\psi_2$. The space $H$ is then made up of orbits.}
    \label{fig:orbit}
\end{figure}

For a given fixed $\psi\in H$, we have a naturally induced map $L_\psi:\ G\rightarrow H$. The images of the map $L_\psi$ identify the possible locations to which the initial ket $\psi$ can travel by actions in $G$. For this reason, we call it the orbit of $\psi$ under $G$, and denote it as $\mathcal{O}_\psi=\{\phi\in H|\ \phi=g\psi,\ \exists g\in G\}$. $\mathcal{O}_\psi$ is exactly the SLOCC-equivalent class that contains the ket $\psi$, which is defined in \cite{dur2000three}. See Fig. \ref{fig:orbit} for illustrations.

The smooth map $L_\psi$ naturally induces a ``push-forward'' map $L_{\psi*}$ from the Lie algebra of $G$ to the tangent space at the point $\psi\in H$: $L_{\psi*}:\ \mathfrak{g}\rightarrow T_\psi(H)$. The elements in $\mathfrak{g}$, usually denoted as $X$, can also be recognized as $2^n$ by $2^n$ matrices. $\forall X\in\mathfrak{g}$, we have a one parameter subgroup $g(t)=\exp(tX)\in G$. $X$ can then be denoted as the tangent of the line $g(t)$, or $X=\big(d/dt\big)|_{t=0}\ g(t)$. With some basic properties of Lie algebra, one achieves:
\begin{equation}
    \begin{split}
        L_{\psi*}(X)=L_{\psi*}\left(\left.\dfrac{d}{dt}\right|_{t=0}g(t)\right)=\left.\dfrac{d}{dt}\right|_{t=0}L_\psi(g(t))=\left.\dfrac{d}{dt}\right|_{t=0}\exp(tX)\psi=X\psi.
    \end{split}
\end{equation}
The first equality through the third equality can be interpreted as: the image of tangent is the tangent of image. The last equality is due to the isomorphism of an Euclidean space with its tangent space. What we have now is $L_{\psi*}(X)=X\psi$. One usually interprets the result as the action of the Lie algebra element $X$ moves the initial ket $\psi$ to $X\psi$. This action is usually termed by physicists as the ``infinitesimal action'' of the group element $\exp(tX)\in G$, when $t$ tends to 0. Mathematically, the ``infinitesimal action'' of a Lie group is the image of push forward of its Lie algebra. The reason why physicists love infinitesimal actions, is that they correspond to the Lie algebra, and Lie algebra has simple structures of linear space.

What can it help us with? The most important question is, what is the dimension of the orbit $\mathcal{O}_\psi$? After some simple derivations, one gets
\begin{equation}\label{dimh}
    \dim(\mathcal{O}_\psi)=\rank(L_\psi)=\rank(L_{\psi*})=\rank(\{X_k\psi\}),\ \text{with}\ \text{span}(X_k)=\mathfrak{g}.
\end{equation}
Here, $X_k$ is a set of bases for the linear space $\mathfrak{g}$.

If the group $G$ is chosen as $G=$GL(2,$\C$)$^{\otimes n}$, the corresponding Lie algebra is the linear space $\mathfrak{g}=\mathfrak{gl}(2,\C)^{\oplus n}$. A generic element in $\mathfrak{gl}(2,\mathbb{C})^{\oplus n}$ is denoted as the superposition
\begin{equation}\label{liealgebra}
    \begin{split}
    &X=\sum_{k=1}^nI_1\otimes\cdots \otimes I_{k-1}\otimes X_k\otimes I_{k+1}\otimes\cdots \otimes I_n,\\
        &\text{where}\ X_k=\begin{pmatrix}
            (q_k+w_k)+i(e_k+t_k) & (u_k+r_k)+i(s_k+v_k),\\
    (u_k-r_k)+i(s_k-v_k) & (q_k-w_k)+i(e_k-t_k)
        \end{pmatrix}\in\mathfrak{gl}(2,\mathbb{C})=M_n(2,\C),\\
        &\text{with}\ r_k,s_k,t_k,u_k,v_k,w_k,e_k,q_k\in\mathbb{R},
    \end{split}
\end{equation}

and $I_k=Q_k$ is the $2\times 2$ identity matrix. We further define the eight independent matrices
\begin{equation}
\begin{split}
    &R_k=\begin{pmatrix}
    0 & 1\\
    -1 & 0
    \end{pmatrix}\quad 
    S_k=\begin{pmatrix}
    0 & i\\
    i & 0
    \end{pmatrix}\quad 
    T_k=\begin{pmatrix}
    i & 0\\
    0 & -i
    \end{pmatrix}\quad 
    E_k=\begin{pmatrix}
    i & 0\\
    0 & i
    \end{pmatrix}\\[1em]
    &U_k=\begin{pmatrix}
    0 & 1\\
    1 & 0
    \end{pmatrix}\quad 
    V_k=\begin{pmatrix}
    0 & i\\
    -i & 0
    \end{pmatrix}\quad 
    W_k=\begin{pmatrix}
    1 & 0\\
    0 & -1
    \end{pmatrix}\quad 
    Q_k=\begin{pmatrix}
    1 & 0\\
    0 & 1
    \end{pmatrix}.
\end{split}
\end{equation}
By adding the subscript $k$ to $X_k$ to denote the qubit which the matrices act on, we can neglect all the identity matrices $I_{i\neq k}$ that are acting on other qubits in \eqref{liealgebra}. Also, we have
\begin{equation}
    X_k=r_kR_k+s_kS_k+t_kT_k+e_kE_k+u_kU_k+v_kV_k+w_kW_k+q_kQ_k.
\end{equation}
Then a Lie algebra element $X\in\mathfrak{gl}(2,\C)$ can be denoted by an $8n$ dimensional vector 
\begin{equation}
X=\sum_{k=1}^nX_k\longmapsto(\{r_k,s_k,t_k,e_k,u_k,v_k,w_k,q_k\}). 
\end{equation}
According to Eq. \eqref{dimh}, we have, when the group is GL(2,$\C$)$^{\otimes n}$,
\begin{equation}
    \dim(\mathcal{O}_\psi^\text{GL(2,$\C$)})=\rank(\{R_k\psi,S_k\psi,T_k\psi,E_k\psi,U_k\psi,V_k\psi,W_k\psi,Q_k\psi\}).
\end{equation}

\subsection{Dimension of orbits as an entanglement witness}
If alternatively, the group we choose for SLOCC-equivalence is SL(2,$\C$)$^{\otimes n}$, one can easily check that the basis vectors of the Lie algebra $X_k$ in \eqref{liealgebra} are traceless $2\times 2$ complex matrices and thus $q_k=e_k=0$. So the global phase and the global amplitude terms are missing for the group SL(2,$\C$)$^{\otimes n}$. In this case, the dimension of the orbit is 
\begin{equation}
    \dim(\mathcal{O}_\psi^\text{SL(2,$\C$)})=\rank(\{R_k\psi,S_k\psi,T_k\psi,U_k\psi,V_k\psi,W_k\psi\}).
\end{equation}

As an example, in one-qubit case, the dimension of the orbit is given by
\begin{equation}
    \begin{array}{cc}
    \hline\hline
       \text{Group} & \dim(\mathcal{O}_\psi) \\
         \hline
        \text{GL}(2,\C)& 4\\
        \hline
        \text{SL}(2,\C)& 4\\
        \hline\hline
    \end{array}
\end{equation}
It can be seen that the orbits by GL(2,$\C$) and SL(2,$\C$) are the same in the ket space. The global phase and global amplitude in GL(2,$\C$) do not bring new information.

For the two-qubit case, a distinction can be found between disentangled and entangled states. The dimension of the orbit is given by
\begin{equation}\label{table2}
    \begin{array}{ccc}
    \hline\hline
    \text{Group}& \multicolumn{2}{c}{\dim(\mathcal{O}_\psi)}\\
    \hline
         & \text{Disentangled} & \text{Entangled} \\
         \hline
        \text{GL}(2,\C) & 6 & 8 \\
        \hline
        \text{SL}(2,\C) & 6 & 6 \\
        \hline\hline
    \end{array}.
\end{equation}
It is interesting to point out that the existence of entanglement between the two qubits prevents the group SL(2,$\C$) from bringing the global phase and global amplitude, and thus $\dim(\mathcal{O}_\psi^\text{GL(2,$\C$)})-\dim(\mathcal{O}_\psi^\text{SL(2,$\C$)})=2$ for entangled two qubit states. However for disentangled states, $\dim(\mathcal{O}_\psi^\text{GL(2,$\C$)})-\dim(\mathcal{O}_\psi^\text{SL(2,$\C$)})=0$, the global phase and global amplitude is preserved by the group SL(2,$\C$). If we define the quantity $\mathcal{W}_1=\dim(\mathcal{O}_\psi^\text{GL(2,$\C$)})-\dim(\mathcal{O}_\psi^\text{SL(2,$\C$)})$ as an entanglement witness, it can successfully detect any entangled states when taking the value 2, while giving 0 for disentangled states in any two-qubit systems.

For the three-qubit case, we have
\begin{equation}\label{table3}
    \begin{array}{ccccc}
    \hline\hline
    \text{Group}& \multicolumn{4}{c}{\dim(\mathcal{O}_\psi)}\\
    \hline
         & \text{Product} & \text{Biseparable} & \text{W-class} & \text{GHZ-class}\\
         \hline
        \text{GL}(2,\C) & 8 & 10 & 14 & 16\\
        \hline
        \text{SL}(2,\C) & 8 & 10 & 14 & 14\\
        \hline\hline
    \end{array}.
\end{equation}
The entanglement witness $\mathcal{W}_1$ now can detect GHZ-type genuine tripartite entanglement, the same as the 3-tangle in \cite{coffman2000distributed}.\\

{\bf Side remarks.} If one wants to study instead LU-equivalence, one can choose the group $G$ to be U(2)$^{\otimes n}$, then
\begin{equation}
    \dim(\mathcal{O}_\psi^\text{U(2)})=\rank(\{R_k\psi,S_k\psi,T_k\psi,E_k\psi\}).
\end{equation}
Or if the group is instead SU(2)$^{\otimes n}$, then
\begin{equation}
    \dim(\mathcal{O}_\psi^\text{SU(2)})=\rank(\{R_k\psi,S_k\psi,T_k\psi\}).
\end{equation}
The situation reduces back to case discussed in \cite{lyons2005minimum}.

\subsection{Orbits in state space}
The orbits in state space is a bit more complicated. We denote an arbitrary state as $[\psi]\in HP$. The smooth action of $G=\ $GL(2,$\C$) on the state space is then a map
\begin{equation}
\begin{split}
    &\sigma:G\times HP\longrightarrow HP,\\
    &\sigma:g\times [\psi]\longmapsto [g\psi]\equiv\pi(g\pi^{-1}([\psi])),\ s.t.\\
    (a)\ &\forall g\in G,\ \sigma_g:HP\rightarrow HP\ \text{is a diffeomorphism}.\\
    (b)\ &\forall g,h\in G,\ \sigma_{gh}=\sigma_g\circ \sigma_h.
\end{split}
\end{equation}
Here $\pi^{-1}([\psi])$ is any representative of $[\psi]$. $\forall\ [\psi]\in HP$, we have a map $\sigma_{[\psi]}:G\rightarrow HP$. The orbit is defined then as $\mathcal{O}_{[\psi]}=\{x\in HP|\ x=g\times [\psi],\ \exists\ g\in G\}$, which is just $\mathcal{O}_{[\psi]}=\text{Im}(\sigma_{[\psi]})$, where Im is the image. This is clearly the SLOCC-equivalent class of the state $[\psi]\in HP$ in the state space. It naturally follows that
\begin{equation}\label{dimox}
    \dim(\mathcal{O}_{[\psi]})=\dim(\text{Im}(\sigma_{[\psi]}))=\dim(G)-\dim(\text{ker}(\sigma_{[\psi]}))=\dim (G)-\dim ({I}_{[\psi]}),
\end{equation}
where the kernel $I_{[\psi]}=\{g\in G|\ g\cdot [\psi]=[\psi]\}$ is the isotropy subgroup of the state $[\psi]$. The dimension $G$ is known. The only question is the dimension of the isotropy subgroup $I_{[\psi]}$. We denote the corresponding isotropy Lie subalgebra as $\mathfrak{i}_{[\psi]}\subset\mathfrak{g}$.

\textbf{Claim 1}. {\it $\forall X\in\mathfrak{g}$, $X\in \mathfrak{i}_{[\psi]}$ iff\ $\exists \lambda\in\mathbb{C}$ s.t. $X\psi=\lambda\psi$, $\psi=\pi^{-1}([\psi])$ is a representative of $[\psi]$.}
\begin{proof}
$X\in \mathfrak{i}_{[\psi]}$ $\Longleftrightarrow$ $\exp(tX)\in I_{[\psi]}$ $\Longleftrightarrow$ $\exp(tX)\cdot [\psi]=[\psi]$ $\Longleftrightarrow$ $\exp(tX)\cdot \psi=\Lambda(t)\psi,\ \exists\ \Lambda$ an arbitrary complex function on $t$ $\Longleftrightarrow$ $X\psi=\Lambda'(0)\psi$, with $\Lambda'(0)$ an arbitrary complex number.
\end{proof}

Suppose $X\in\mathfrak{i}_{[\psi]}$, according to Claim 1, $X\psi=\lambda\psi$ for some complex $\lambda=\lambda_1+i\lambda_2$. Writing this in a more explicit form
\begin{equation}
    \sum_k(r_kR_k|\psi\rangle+s_kS_k|\psi\rangle+t_kT_k|\psi]\rangle+e_kE_k|\psi\rangle+u_kU_k|\psi\rangle+v_kV_k|\psi\rangle+w_kW_k|\psi\rangle+q_kQ_k|\psi\rangle)=\lambda_1|\psi\rangle+\lambda_2i|\psi\rangle.
\end{equation}
By defining the matrix
\begin{equation}
    M=(\{R_k|\psi\rangle,S_k|\psi\rangle,T_k|\psi\rangle,E_k|\psi\rangle,U_k|\psi\rangle,V_k|\psi\rangle,W_k|\psi\rangle,Q_k|\psi\rangle\},-|\psi\rangle,-i|\psi\rangle),
\end{equation}
it is a linear transformation from $\R^{8n+2}$ to $\R^{2^{n+1}}$. For any given $X\in\R^{8n}$ in $\mathfrak{gl}(2,\C)$, if $X\in\mathfrak{i}_{[\psi]}$, we must have one unique vector $v=(X^T,\lambda_1,\lambda_2)^T$ such that $Mv=0$. On the contrary, if the vector $v=(X^T,\lambda_1,\lambda_2)^T$ is in the kernel of $M$, then $X|\psi\rangle=(\lambda_1+i\lambda_2)|\psi\rangle$ so $X\in\mathfrak{i}_x$. Thus we have a one-to-one and onto map from $\mathfrak{i}_{[\psi]}$ to $\ker(M)$. So $\dim(\mathfrak{i}_{[\psi]})=\dim(\ker(M))$. According to \eqref{dimox}, we have
\begin{equation}
\begin{split}
    \dim(\mathcal{O}_{[\psi]})=&\dim(G)-\dim(I_{[\psi]})=\dim(G)-\dim(\mathfrak{i}_{[\psi]})=\dim(G)-\dim(\ker(M))\\
    =&8n-(8n+2-\rank(M))=\rank(M)-2.
    \end{split}
\end{equation}
This is a generalization of the work in \cite{lyons2005minimum}.

It can be found that $E_k|\psi\rangle=i|\psi\rangle$ and $Q_k|\psi\rangle=|\psi\rangle$, so the last two columns do not add new ranks to $M$, and we have $\rank (M)=\rank (X\psi)=\dim(\mathcal{O}_\psi)$.
So $\dim(\mathcal O_{[\psi]}^\text{GL(2,$\C$)})=\dim(\mathcal O_\psi^\text{GL(2,$\C$)})-2$.

If the group we used from the very beginning is instead SL(2,$\C$)$^{\otimes n}$, we still have $\dim(\mathcal O_{[\psi]}^\text{SL(2,$\C$)})=\rank(M)-2$, but $M$ is now a linear transformation from $\R^{6n+2}$ to $\R^{2^{n+1}}$. Since the last two columns in M compensate the $2n$ missing columns, we have $\dim(\mathcal O_{[\psi]}^\text{GL(2,$\C$)})=\dim(\mathcal O_{[\psi]}^\text{SL(2,$\C$)})$. So the orbits in the state space are the same by GL(2,$\C$) and SL(2,$\C$), different from the case in the ket space. This is the justification of using SL(2,$\C$) instead of GL(2,$\C$) when the situation is focused on the state space.

\begin{figure}[b]
    \centering
    \includegraphics[width=0.8\textwidth]{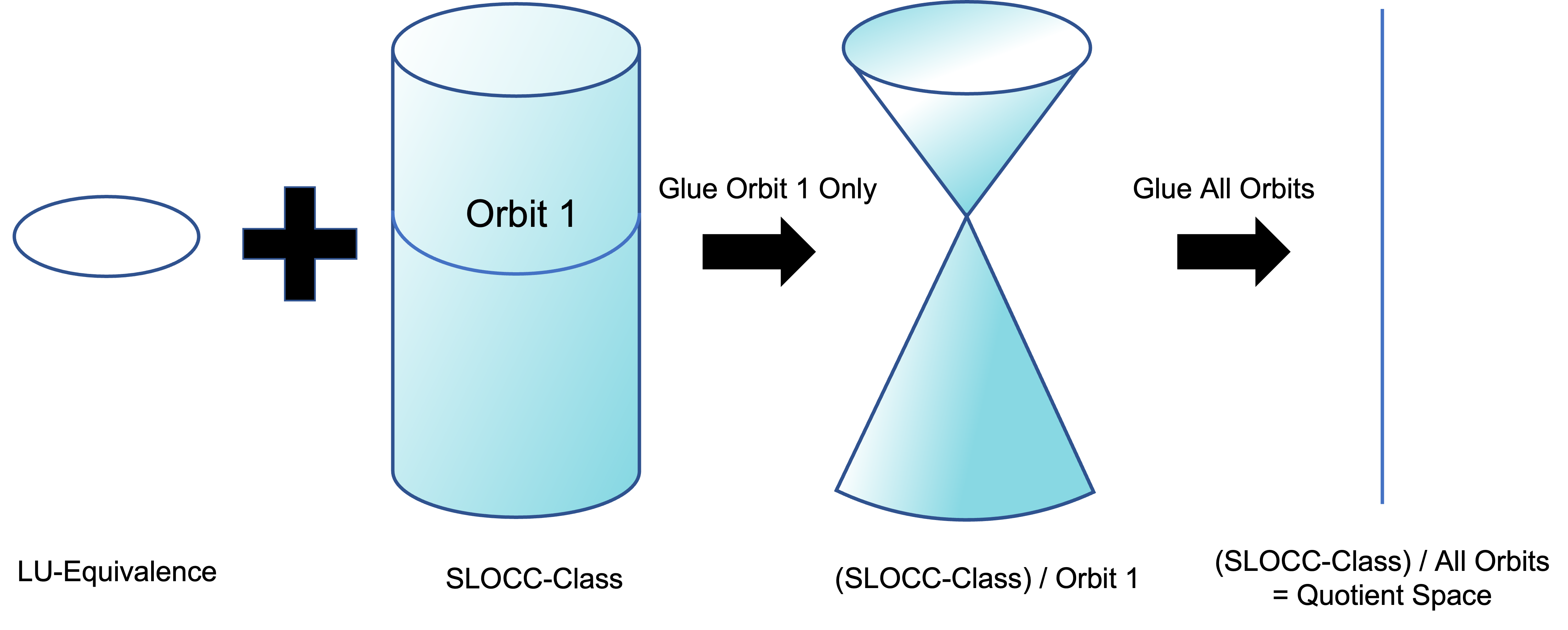}
    \caption{The illustration of how gluing of orbits can transform an original space into a quotient space.}
    \label{fig:quotient}
\end{figure}

\section{Quotient Space}
In the Hilbert space, if two kets differ by only a complex number $C$, one considers that they are the same state. One then glues all these kets together and identifies them as the same point. The resulting space is the state space, as we have defined earlier. 

Similarly, entanglement is insensitive of local bases, and two states that are LU-equivalent share the same entanglement. For this reason, one wants to glue all LU-equivalent states together and identifies them as one single point. One denote the resulting space as $\left.\mathcal{O}_{[\psi]}^\text{GL(2,$\C$)}\right/\text{U(2)}^{\otimes n}$, and it is called a quotient space. It does not distinguish states related by LU equivalence within the SLOCC class. See Fig. \ref{fig:quotient} for illustrations. 

According to D\"ur, Vidal, and Cirac \cite{dur2000three}, the GHZ SLOCC-class has 5 free parameters, and the $W$ SLOCC-class has 3 free parameters. Here the number of ``free parameter'' means the dimension of the quotient space that we have just defined. Generically, given an arbitrary state, how do we find out its ``free parameter''? We have a nice theorem to attack the question \cite{bredon1972introduction}, given by:\\

\noindent{\bf Principal Orbit Theorem.} {\it G is a compact Lie group acting isometrically on a manifold M. Then there exists a unique maximal orbit type, the union of which is open and dense.}\\

The proof is found in the reference book \cite{bredon1972introduction}. As for our case, the group $G=\text{U(2)}^{\otimes n}$ is a compact Lie group. The manifold $M=\mathcal{O}_{[\psi]}^\text{GL(2,$\C$)}$ is an SLOCC-class. Principal orbits are those with the largest dimension. The union of principal orbits is dense, and hence occupies almost everywhere in the SLOCC orbit. Then the dimension of our quotient space is given by $\dim\left(\left.\mathcal{O}_{[\psi]}^\text{GL(2,$\C$)}\right/\text{U(2)}^{\otimes n}\right)=\dim\left(\mathcal{O}_{[\psi]}^\text{GL(2,$\C$)}\right)-\dim \text{of a U(2)}^{\otimes n}$-principal orbit.

The dimension of $\mathcal{O}_{[\psi]}^\text{GL(2,$\C$)}$ can be determined using the method from the previous section. The dimension of a U(2) principal orbit can be determined by applying the Monte Carlo random number generator. Since the principal orbit union is dense, a randomly generated element in GL(2,$\C$)$^{\otimes n}$ will almost always bring the state $[\psi]$ to a state $[\widetilde\psi]$ in a U(2) principal orbit. The dimension of the U(2) principal orbit is then $\dim\left(\mathcal{O}_{[\widetilde\psi]}^\text{U(2)}\right)$.

For short notations, we denote $D_1=\dim\left(\mathcal{O}_{[\psi]}^\text{GL(2,$\C$)}\right)$, the dimension of the SLOCC equivalent class; $D_2=\dim\left(\mathcal{O}_{[\widetilde\psi]}^\text{U(2)}\right)$, the dimension of the principal U(2) orbit; and $D_3=D_1-D_2=\dim\left(\left.\mathcal{O}_{[\psi]}^\text{GL(2,$\C$)}\right/\text{U(2)}^{\otimes n}\right)$, the dimension of the quotient space.

\subsection{One-Qubit Case}
For an arbitrary one-qubit state, $D_1=2$, $D_2=2$, and $D_3=0$. It is a condensed single point. All one-qubit states are in the same SLOCC class and the same LU class.

\subsection{Two-Qubit Case}
There are two SLOCC-equivalent class: disentangled class and entangled class.
For the entangled class, $D_1=6$, $D_2=5$ so $D_3=1$, which corresponds exactly to the well-known Schmidt decomposition:
\begin{equation}
    |\psi\rangle=\cos\theta|00\rangle+\sin\theta|11\rangle,
\end{equation}
with one free real parameter $\theta\in(0,\pi/4]$.

For the disentangled class, $D_1=4$, $D_2=4$, so $D_3=0$, corresponding to the ending point of the Schmidt decomposition, $\theta=0$.

A detailed summary can be found in Table \ref{twoqubittable}.
\begin{table}[h]
    \centering
    \begin{tabular}{c cccc}
    \hline 
    \hline \\[-0.5em]
        SLOCC class & Representative &$D_1$&$D_2$&$D_3$ \\[0.5em]
        \hline\\[-0.5em]
        Entangled & $|00\rangle+|11\rangle$&6&5&1\\[0.5em]
        \hline\\[-0.5em]
        Disentangled &$|00\rangle$&4&4&0\\[0.5em]
        \hline 
        \hline
    \end{tabular}
    \caption{Dimensions for two-qubit case.}
    \label{twoqubittable}
\end{table}

\subsection{Three-Qubit Case}
There are six SLOCC-equivalent class according to \cite{dur2000three}.

For the product class, $D_1=6$ and $D_2=6$, so $D_3=0$. Again, this is a single point. 

For the biseparable class, $D_1=8$ and $D_2=7$, so $D_3=1$. This corresponds to one single qubit composing an entangled two-qubit pair, which can in turn be identified with the Schmidt decomposition as well.

For the $W$ class, $D_1=12$, $D_2=9$, so $D_3=3$, matching the results in \cite{dur2000three}, where three free parameters are needed in $W$ class:
\begin{equation}
    |\psi\rangle_\text{W}=a|000\rangle+b|100\rangle+c|010\rangle+d|001\rangle,
\end{equation}
with $a^2+b^2+c^2+d^2=1$.

For the GHZ class, $D_1=14$, $D_2=9$, so $D_3=5$, matching the results in \cite{dur2000three}, where five free parameters are needed in GHZ class:
\begin{equation}
\begin{split}
    |\psi\rangle_\text{GHZ}=&\cos\delta|000\rangle+\sin\delta\ e^{i\phi}|\phi_\alpha\rangle|\phi_\beta\rangle|\phi_\gamma\rangle\\
    =&\cos\delta|000\rangle+\sin\delta\ e^{i\phi}(\cos\alpha|0\rangle+\sin\alpha|1\rangle)\otimes(\cos\beta|0\rangle+\sin\beta|1\rangle)\otimes(\cos\gamma|0\rangle+\sin\gamma|1\rangle).
\end{split}
\end{equation}

A detailed summary can be found in Table \ref{threequbittable}.
\begin{table}[h]
    \centering
    \begin{tabular}{c cccc}
    \hline 
    \hline \\[-0.5em]
        SLOCC class & Representative &$D_1$&$D_2$&$D_3$ \\[0.5em]
        \hline\\[-0.5em]
        GHZ & $|000\rangle+|111\rangle$&14&9&5\\[0.5em]
        \hline\\[-0.5em]
        $W$ &$|100\rangle+|010\rangle+|001\rangle$&12&9&3\\[0.5em]
        \hline\\[-0.5em]
        Biseparable&$|000\rangle+|011\rangle$&8&7&1\\[0.5em]
        \hline\\[-0.5em]
        Product&$|000\rangle$&6&6&0\\[0.5em]
        \hline 
        \hline
    \end{tabular}
    \caption{Dimensions for two-qubit case.}
    \label{threequbittable}
\end{table}

\begin{table}[h]
\centering
        \begin{tabular}{ccccc}
        \hline\hline\\[-0.5em]
            \text{SLOCC class}& \text{Representative} & $D_1$ & $D_2$ & $D_3$\\[0.5em]
            \hline\\[-0.9em]
            \text{GHZ} & $|0000\rangle+|1111\rangle$ & 18 & 12 & 6 \\[0.1em]
            \hline\\[-0.9em]
            $W$ & $|1000\rangle+|0100\rangle+|0010\rangle+|0001\rangle$ & 16 & 12 & 4\\[0.1em]
            \hline\\[-0.9em]
            $C_4$ & $|0011\rangle+|1100\rangle+|0101\rangle+|1010\rangle+|0110\rangle+|1001\rangle$ & 22 & 12 & 10\\[0.1em]
            \hline\\[-0.9em]
            $\kappa_4$ & $|0000\rangle+|0011\rangle+|1010\rangle-|1111\rangle$ & 22 & 12 & 10\\[0.1em]
            $E_4$ & $|0000\rangle+|0101\rangle+|1001\rangle-|1111\rangle$ & 22 & 12 & 10\\[0.1em]
            $L_4$ & $|0000\rangle+|0011\rangle+|1001\rangle-|1111\rangle$ & 22 & 12 & 10\\[0.1em]
            \hline\\[-0.9em]
            $H_4$ & $|0011\rangle+|0110\rangle+|1100\rangle$ & 20 & 12 & 8\\[0.1em]
            $\lambda_4$ & $|0101\rangle+|0110\rangle+|1010\rangle$ & 20 & 12 & 8\\[0.1em]
            $M_4$ & $|0011\rangle+|0101\rangle+|1100\rangle$ & 20 & 12 & 8\\[0.1em]
            \hline\\[-0.9em]
            $\pi_4$ & $|0000\rangle+|0011\rangle+|0101\rangle+|0110\rangle+|1010\rangle+|1111\rangle$ & 20 & 12 & 8\\[0.1em]
            $\theta_4$ & $|0000\rangle+|0101\rangle+|0110\rangle+|1010\rangle+|1100\rangle+|1111\rangle$ & 20 & 12 & 8\\[0.1em]
            $\sigma_4$ & $|0000\rangle+|0011\rangle+|1001\rangle+|1010\rangle+|1100\rangle+|1111\rangle$ & 20 & 12 & 8\\[0.1em]
            $\rho_4$ & $|0000\rangle+|0011\rangle+|0110\rangle+|1010\rangle+|1100\rangle+|1111\rangle$ & 20 & 12 & 8\\[0.1em]
            $\xi_4$ & $|0000\rangle+|0110\rangle+|1001\rangle+|1010\rangle+|1100\rangle+|1111\rangle$ & 20 &  12& 8\\[0.1em]
            $\epsilon_4$ & $|0000\rangle+|0011\rangle+|0110\rangle+|1001\rangle+|1010\rangle+|1111\rangle$ & 20 & 12 & 8\\[0.1em]
            \hline\\[-0.9em]
            $\chi_4$ & $|0000\rangle+|0011\rangle+|0110\rangle+|1010\rangle+|1100\rangle-|1111\rangle$ & 24 & 12 & 12\\[0.1em]
            \hline\\[-0.9em]
            $\psi_4$ & $|0000\rangle+|0101\rangle+|1010\rangle-|1111\rangle$ & 20 & 12 & 8 \\[0.1em]
            $\phi_4$ & $|0000\rangle+|0011\rangle+|1100\rangle-|1111\rangle$ & 20 & 12 & 8\\[0.1em]
            $\mu_4$ & $|0000\rangle+|0110\rangle+|1001\rangle-|1111\rangle$ & 20 & 12 & 8\\[0.1em]
            \hline\\[-0.9em]
            $\varphi_4$ & $|0001\rangle+|0110\rangle+|1011\rangle$ & 18 & 12 & 6\\[0.1em]
            $\vartheta_4$ & $|0010\rangle+|0101\rangle+|1011\rangle$ & 18 & 12 & 6\\[0.1em]
            $\tau_4$ & $|0001\rangle+|0111\rangle+|1010\rangle$ & 18 & 12 & 6\\[0.1em]
            $\varrho_4$ & $|0010\rangle+|0111\rangle+|1001\rangle$ & 18 & 12 & 6\\[0.1em]
            \hline\\[-0.9em]
            $\zeta_4$ & $|0000\rangle+|1011\rangle+|1100\rangle$ & 18 & 12 & 6\\[0.1em]
            $\iota_4$ & $|0000\rangle+|0011\rangle+|1101\rangle$ & 18 & 12 & 6\\[0.1em]
            \hline\\[-0.9em]
            $\nu_4$ & $|0010\rangle+|0101\rangle+|1001\rangle+|1011\rangle$ & 20 & 12 & 8\\[0.1em]
            \hline\\[-0.9em]
            $\omega_4$ & $|0000\rangle+|0101\rangle+|1000\rangle+|1110\rangle$ & 20 & 12 & 8\\[0.1em]
            \hline\\[-0.9em]
            $\varpi_4$ & $|0010\rangle+|0101\rangle+|1000\rangle+|1100\rangle$ & 20 & 12 & 8\\[0.1em]
            \hline
            \hline\\[-0.9em]
            $A\mbox{-}B\mbox{-}C\mbox{-}D$ & $|0000\rangle$ & 8 & 8 & 0\\[0.1em]
            \hline\\[-0.9em]
            $A\mbox{-}B\mbox{-}CD$ & $|0000\rangle+|0011\rangle$ & 10 & 9 & 1\\[0.1em]
            \hline\\[-0.9em]
            $AB\mbox{-}CD$ & $|0000\rangle+|0011\rangle+|1100\rangle+|1111\rangle$ & 12 & 10 & 2\\[0.1em]
            \hline\\[-0.9em]
            $A\mbox{-}\text{GHZ}$ & $|0000\rangle+|0111\rangle$ & 16 & 11 & 5\\[0.1em]
            \hline\\[-0.9em]
            $A\mbox{-}W$ & $|0100\rangle+|0010\rangle+|0001\rangle$ & 14 & 11 & 3\\[0.1em]
            \hline
            \hline
        \end{tabular}
        \caption{Dimensions for four-qubit case}
        \label{fourqubittable}
\end{table}

\subsection{Four-Qubit Case}
There is no clear classification of SLOCC equivalence for four qubits. But we do have particular well-known states, for which we can apply the above techniques.

The GHZ class is the $\text{GL}(2,\C)^{\otimes n}$-orbit of $|0000\rangle+|1111\rangle$. $D_1=18$, $D_2=12$ so $D_3=6$. We have the following guess for the general form:
\begin{equation}
    \begin{split}
        |\psi\rangle_\text{GHZ}=&\cos\delta|0000\rangle+\sin\delta\ e^{i\phi}|\phi_\alpha\rangle|\phi_\beta\rangle|\phi_\gamma\rangle|\phi_\theta\rangle,
    \end{split}
\end{equation}
which is a generalization of the GHZ class in three qubits, and contains exactly six free parameters.

The $W$ class is the $\text{GL}(2,\C)^{\otimes n}$-orbit of $|1000\rangle+|0100\rangle+|0010\rangle+|0001\rangle$. $D_1=16$, $D_2=12$, so $D_3=4$. A natural guess is then
\begin{equation}
    \begin{split}
        |\psi\rangle_W=a|0000\rangle+b|1000\rangle+c|0100\rangle+d|0010\rangle+e|0001\rangle,
    \end{split}
\end{equation}
with $a^2+b^2+c^2+d^2+e^2=1$. This is a generalization of the $W$ class in three qubits, and contains exactly four free parameters.

The $C_4$ class introduced in \cite{li2006simple}, as the orbit of $|0011\rangle+|1100\rangle+|0110\rangle+|1001\rangle+|0101\rangle+|1010\rangle$.
$D_1=22$, $D_2=12$, so $D_3=10$. This is a larger class than the GHZ class! It is quite surprising that the GHZ class is no more the largest class for four qubits.

The $\phi_4$ class introduced in \cite{briegel2001persistent} as the orbit of $|0000\rangle+|0011\rangle+|1100\rangle-|1111\rangle$. $D_1=20$, $D_2=12$, so $D_3=8$.

In \cite{li2007classification}, 28 distinct genuinely entangled SLOCC equivalent classes are given. A detailed summary of the dimensions for all these 28 classes, together with all the non-genuinely entangled states, can be found in Table \ref{fourqubittable}.

One is eager to know the largest SLOCC class for four qubits. This can be done by setting $|\psi\rangle=(c_1,c_2,c_3,c_4,c_5,c_6,c_7,c_8)$ as a generic quantum ket and calculate the dimension of $\mathcal{O}_{[\psi]}^{\text{GL(2,$\C$)}}$, whose value is $D_1=24$, the same as that of the class $\chi_4$. However, the dimension of the state space is $\dim\left(\C P^{15}\right)=30$. One can see that the dimension of the largest SLOCC orbit is much smaller than the whole state space, and is thus of zero measure. Thus for four qubits there are infinitely many SLOCC classes, and one needs multiple continuous variables to represent all those classes. This is a confirmation of the statement in \cite{dur2000three} with a rigorous proof.

\subsection{Dimension of orbits as another entanglement witness}

By direct observation, $D_2$ is also an entanglement witness. It can strictly distinguish different types of entanglement. For example, for two qubits, entangled states have $D_2=5$ and disentangled states have $D_2=4$. For three qubits, genuinely entangled states have $D_2=9$. Biseparable states have $D_2=7$. Product states have $D_2=6$. For four qubits, genuinely entangled states have $D_2=12$. $A\mbox{-}BCD$ type states have $D_2=11$. $AB\mbox{-}CD$ type states have $D_2=10$. $A\mbox{-}B\mbox{-}CD$ type states have $D_2=9$. $A\mbox{-}B\mbox{-}C\mbox{-}D$ type states have $D_2=8$. Thus the value $D_2$ can be seen as an entanglement witness, and quantify the ``size'' of the entanglement type. The larger $D_2$ is, the ``larger'' the entanglement type is.

\section{Summary}
In this work, we have studied the orbit structure of two different groups, GL(2,$\C$)$^{\otimes n}$ and SL(2,$\C$)$^{\otimes n}$ acting on the Hilbert space and state space, respectively. It turns out the two groups make no difference for the state space, which distinguishes the overall complex factor. However, the dimensions of the orbits by the two groups on the Hilbert space are different. Surprising, we found the the difference of the dimensions can be considered as an entanglement witness. 

Next, we treat each SLOCC-class as the quotient space of the state space over SLOCC-equivalence. We developed a tool to determine the dimension of each SLOCC-class. For one-qubit, two-qubit, and three-qubit cases, our results are fully in agreement with previous discoveries. Specifically, in three-qubit case, the number of free parameters for the GHZ class and the $W$ class are identified. In four-qubit case, we find that the class containing the GHZ state, which we call GHZ class, is no longer the largest class. Furthermore, it is found that unlike the case of three qubits where there are only 6 SLOCC-classes, the number of SLOCC class for four qubits is infinite. We also find an interesting quantity $D_2$, which can distinguish types of entanglement in a multi-qubit system.

\section{Acknowledgements}
The author thanks Prof. J.H. Eberly for valuable discussions and continuous assistance. Financial support was provided by National Science Foundation grants PHY-1501589 and PHY-1539859 (INSPIRE).

\bibliography{dimension}
\end{document}